# Axial Attention Transformer Networks: A New Frontier in Breast Cancer Detection

1st Weijie He
University of California, Los Angeles
Los Angeles, USA

2nd Runyuan Bao
Johns Hopkins University
Baltimore, USA

3rd Yiru Cang
Northeastern University
Boston, USA

4th Jianjun Wei
Washington University in St. Louis
St Louis, USA

5th Yang Zhang
Boston University
Boston, USA

6th Jiacheng Hu*
Tulane University
New Orleans, USA

*Abstract*—This paper delves into the challenges and advancements in the field of medical image segmentation, particularly focusing on breast cancer diagnosis. The authors propose a novel Transformer-based segmentation model that addresses the limitations of traditional convolutional neural networks (CNNs), such as U-Net, in accurately localizing and segmenting small lesions within breast cancer images. The model introduces an axial attention mechanism to enhance the computational efficiency and address the issue of global contextual information that is often overlooked by CNNs. Additionally, the paper discusses improvements tailored to the small dataset challenge, including the incorporation of relative position information and a gated axial attention mechanism to refine the model's focus on relevant features. The proposed model aims to significantly improve the segmentation accuracy of breast cancer images, offering a more efficient and effective tool for computer-aided diagnosis.

*Keywords-Medical Image Segmentation, Deep Learning, Breast Cancer Diagnosis, Axial Attention Mechanism*

## I. Introduction

The semantic segmentation of medical images is extremely important in the process of computer-aided diagnosis, as accurate localization and segmentation of lesions can provide strong assistance to doctors in clinical diagnosis. With the continuous improvement of computing power, a large number of researches related to deep learning have emerged in recent years, and many deep learning-based medical image segmentation algorithms have been successful. Although deep learning has made significant progress in the field of medical image analysis, it still faces many difficulties and challenges.

Currently, there are excellent segmentation networks such as U-Net [1]. However, U-Net and other convolutional neural network-based segmentation models [2] have not fully met the strict requirements for segmentation accuracy in medical image analysis. In the task of breast cancer segmentation, the lesions usually account for a very small proportion of the entire breast cancer image, and most of the features extracted by the segmentation model are useless background information. In addition, considering the interference of occlusions between different tissues in the breast cancer segmentation task, how to accurately locate the lesions remains a challenging research work.

The size of the convolutional kernel is fixed. Due to the inherent limitations of convolutional operations, each convolutional kernel only focuses on a local set of pixels in the entire image, forcing the network to pay more attention to local information rather than global contextual information, resulting in the lack of ability of convolutional neural networks to learn global information. Although subsequent research has proposed some methods [3, 4] to compensate for the shortcomings of convolutional neural networks, these methods are ultimately only a temporary solution and do not fundamentally solve the problems faced by convolutional neural networks. The summary of the second chapter has listed the advantages of Transformers [5] compared to convolutional neural networks, and Transformers are known for their ability to effectively model the global dependencies between inputs and outputs. However, most previous image segmentation methods have not been optimized for breast cancer segmentation tasks, and there is still a large room for improvement in lesion segmentation accuracy, so this paper focuses on the application of Transformers in breast cancer image segmentation.

This paper first makes improvements to the problem of difficult localization of small lesions in breast cancer images, mainly including:

(1) In response to the large size of medical images, the encoder structure of the Transformer is simplified, and the attention layer is optimized, replacing the traditional self-attention mechanism with an axial attention mechanism. This reduces the computational complexity of the Transformer and improves the computational efficiency of the model on breast cancer images.

(2) In response to the small size of medical image datasets, where the Transformer's segmentation performance is poor due to the lack of inductive bias on small datasets, this chapter adds relative position information during the calculation process of the axial attention mechanism.

(3) In order to further enhance the key information in the relative position encoding and suppress the interference caused by inaccurate position information, this chapter proposes an axial attention mechanism with gated units.

(4) By using the axial attention module with added gating units and relative position encoding to replace the traditional self-attention module, a Transformer segmentation model based on axial attention is constructed.

## II. RELATED WORK

In clinical diagnosis, doctors also need to analyze the entire image from a global perspective to make judgments. Obviously, the Transformer is closer to the actual diagnostic process of doctors, so the potential of the Transformer-based segmentation network in medical image segmentation tasks is very large. However, the Transformer segmentation model based on the traditional attention mechanism is not without shortcomings. It can be seen that the huge computational cost of the traditional Transformer limits its potential in the analysis of large-sized images (such as medical images). In response to the above situation, Wang et al. [6] proposed Axial-Attention and probabilistic methods, which decomposes the two-dimensional self-attention mechanism into two one-dimensional self-attention mechanisms. This can not only integrate global attention information and model long-distance spatial dependencies but also effectively reduce the computational cost of the attention mechanism, making it possible for the Transformer-based segmentation model to handle large-scale medical images.

In recent years, deep learning has revolutionized the field of medical image analysis, particularly in cancer diagnosis and segmentation. The application of convolutional neural networks (CNNs) has been instrumental in many image classification and segmentation tasks. Xiao et al. [7] demonstrated the effectiveness of CNNs in classifying breast cancer cytopathology images, providing a foundational approach for image-based cancer detection. However, despite their success, CNN-based models like U-Net struggle with accurately segmenting small lesions due to their inability to capture global context efficiently. This limitation has motivated the exploration of alternative architectures, such as attention-based models and Transformers.

The U-Net architecture has been a standard in medical image segmentation, yet it falls short in scenarios requiring high precision for small lesion detection, as in breast cancer images. Zhu et al. [8] proposed Attention-U-Net, which introduced an attention mechanism into the U-Net framework to improve segmentation accuracy by focusing on relevant regions of the image. Their work laid the groundwork for integrating attention mechanisms into segmentation networks, which directly inspired the axial attention mechanism introduced in this study. Zhu et al.'s work emphasized the importance of selectively enhancing the attention on critical areas within medical images, particularly where lesions are small and hard to distinguish. Xu et al. [9] highlighted the importance of data preprocessing and optimization techniques in improving deep learning performance in medical diagnostics, especially when dealing with limited datasets. Their emphasis on the preprocessing phase aligns with this study's strategy of using relative position encoding to enhance the performance of the Transformer model on small datasets.

Yan et al. [10] extended the use of deep learning in cancer-related tasks beyond image segmentation by applying neural networks to survival prediction across diverse cancer types. Their work demonstrates the versatility of deep learning in medical applications, reinforcing the potential of neural networks to model complex medical data. Although survival prediction is not directly related to image segmentation, the optimization techniques and neural network structures they explored contribute to the broader understanding of how deep learning can be adapted to various medical challenges, including segmentation. Gao et al. [11] explored the optimization of neural networks through graph-based techniques, which, while not specific to medical imaging, provide insights into enhancing network efficiency and accuracy. Their work on optimizing text classification models has parallels in this paper's use of the axial attention mechanism to optimize the efficiency of the Transformer model, particularly in high-resolution breast cancer image segmentation. Similarly, Yang et al. [12] developed dynamic neural network architectures for predicting sequential medical visits, showcasing how advanced neural architectures can be applied to complex medical data. This highlights the broader applicability of flexible neural network models in medical contexts, which informs the current work's approach to developing adaptable and efficient architectures for medical image segmentation. Additionally, Wang et al. [13] employed functional annotations to refine probabilistic models, highlighting methods that can inform more precise lesion segmentation in breast cancer imaging. Zheng et al. [14] proposed novel enhancements to deep learning optimizers, introducing adaptive friction mechanisms with sigmoid and tanh functions. Although their work focuses on optimizer improvements, the general principle of improving convergence and model performance has implications for the optimization strategies used in this study.

## III. BACKGROUND

The gated axial attention mechanism introduced here similarly aims to refine the focus of the model on relevant features, enhancing segmentation accuracy and robustness in challenging datasets with limited size.

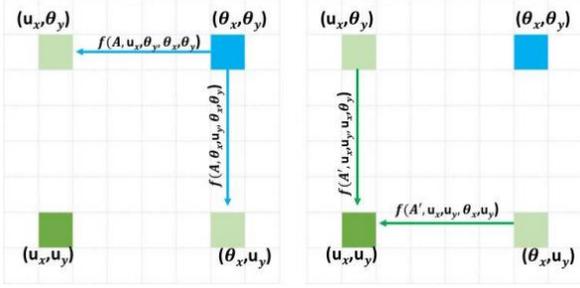

Figure 1. Diagram of the Cross-Attention Module

The principle of the axial attention approach is to repeatedly use the cross-attention module. As shown in Figure 1, if you want to calculate the correlation between a pixel $(\theta_x, \theta_y)$ and any pixel $(u_x, u_y)$ in the figure, you can obtain it through two cross-attentions. Specifically, in the first cross-attention, the blue pixel $(\theta_x, \theta_y)$ interacts with the surrounding light green pixels that have a cross relationship, such as the points along the vertical axis $(\theta_x, u_y)$ and the points along the horizontal axis $(u_x, \theta_y)$. The second cross-attention can be based on the first calculation, and through the cross-relationship of the light green features, establish a relationship breast cancer image segmentation tasks. This section mainly introduces the overall structure of the Transformer segmentation model based on axial attention and the important axial attention module.

*A. Relative Position Information and Axial Attention*

The current situation of small data volume in medical image datasets limits the performance of the Transformer from another aspect. In response to the characteristic of small data volume in medical image datasets, the solution of this paper is to add relative position information to the axial attention mechanism. Relative position information plays a role in supplementing position information in the process of axial attention computation, supplementing information from the perspective of the query, and supplementing information from the perspective of the key. The more information related to the segmentation task is gathered, the more useful information is obtained, and the more accurate the segmentation result is.

Taking Figure 3 as an example, assuming the input feature map is xx, the height of the feature map is HH, the width is WW, and the number of channels is CC. The traditional self-attention mechanism can be expressed as follows:

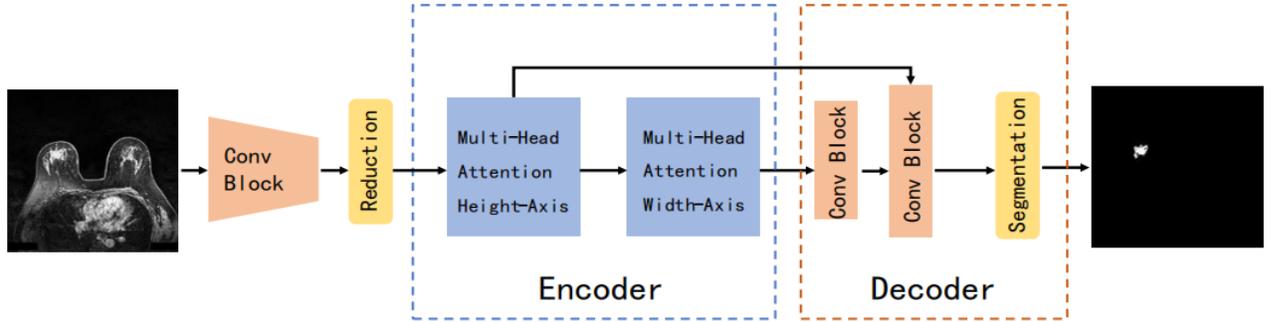

Figure 2. Structural diagram of the Transformer segmentation model based on axial attention.

with the deep green target pixel $(u_x, u_y)$. In the above example, the light green pixels $(u_x, \theta_y)$ and $(\theta_x, u_y)$ have already included the dependency relationship with the blue pixel $(\theta_x, \theta_y)$ in the first information interaction, so any two points in the two-dimensional image only need two cross-attentions to establish information interaction. Compared with the traditional attention mechanism, the axial attention mechanism can greatly reduce the network's computational load to O($hw$), while still effectively achieving the interaction of global contextual information.

IV. METHOD

Traditional convolutional neural networks lack the ability to model long-range dependencies in images, while Transformers are effective in modeling global dependencies. This chapter focuses on exploring the application of Transformers in the field of medical image segmentation and proposes a solution to the problem of locating small lesions in

$$y_{ij} = \sum_{h=1}^{H} \sum_{w=1}^{W} \text{softmax}(q_{ij}^T k_{hw}) v_{hw}$$

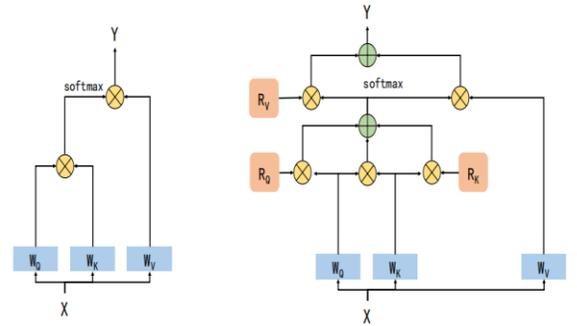

Figure 3. The schematic diagram of the attention mechanism with the addition of relative position encoding.

The self-attention mechanism with added relative position encoding can be expressed as follows:

$$y_{ij} = \sum_{h=1}^{H}\sum_{w=1}^{W} \text{softmax}\bigl(q_{ij}^{\text{T}}k_{hw} + q_{ij}^{\text{T}}R_Q + k_{hw}^{\text{T}}R_K\bigr)(v_{hw} + R_V)$$

Equation adds three learnable relative position encodings compared to the traditional attention mechanism. In experiments, the relative position encodings are a set of randomly initialized parameters, that is, learnable positional information. Although it increases the computational cost to some extent, the relative position information compensates for the weak inductive bias modeling ability of the Transformer when the data volume is small, which can be said to further reduce the obstacles of the Transformer in the breast cancer segmentation task by adding relative position encoding.

Assuming it is decomposed, it can be obtained along the height axis attention and along the width axis attention, that is, a two-dimensional attention is decomposed into two one-dimensional axial attentions.

The one-dimensional axial attention along the height can be expressed as follows:

$$y_{ij} = \sum_{h=1}^{H} \text{softmax}\bigl(q_{ij}^{\text{T}}k_{hj} + q_{ij}^{\text{T}}R_Q + k_{hj}^{\text{T}}R_K\bigr)(v_{hj} + R_V)$$

The one-dimensional axial attention along the width can be expressed as follows:

$$y_{ij} = \sum_{w=1}^{W} \text{softmax}\bigl(q_{ij}^{\text{T}}k_{iw} + q_{ij}^{\text{T}}R_Q + k_{iw}^{\text{T}}R_K\bigr)(v_{iw} + R_V)$$

### B. Gated Axial Attention Module

Inspired by Attention U-Net, this paper adds a gating mechanism to the axial attention to further amplify key information and suppress interference items. The motivation for this improvement is that on small datasets such as breast cancer images, positional biases are difficult to learn, and therefore the long-distance information interaction in the Transformer is not always accurate. Although the introduction of relative positional encoding in the previous section has alleviated this issue to some extent, it has overlooked the fact that inaccurate positional encoding may reduce the model's segmentation accuracy

Therefore, based on Equations, this paper adds gate units to form an improved axial attention mechanism, as shown in Figure 4. Gate units can control the influence of positional biases on non-local context encoding and regulate the learning process of relative positional encoding to some extent. The modified axial attention mechanism calculation process can be expressed as follows:

$$y_{ij} = \sum_{h=1}^{H} \text{softmax}\bigl(q_{ij}^{\text{T}}k_{hj} + G_Q q_{ij}^{\text{T}}R_Q + G_K k_{hj}^{\text{T}}R_K\bigr)(v_{hj} + G_V R_V)$$

$$y_{ij} = \sum_{w=1}^{W} \text{softmax}\bigl(q_{ij}^{\text{T}}k_{iw} + G_Q q_{ij}^{\text{T}}R_Q + G_K k_{iw}^{\text{T}}R_K\bigr)(v_{iw} + G_V R_V)$$

In Equations, the three gate units, like relative positional encodings, are learnable parameters. When the data volume of the breast cancer image dataset is not sufficient for the network to learn accurate positional biases, the gate units will give the relative positional encoding a smaller weight to suppress inaccurate relative positional information and avoid interference with the model's segmentation accuracy. Conversely, the gate units give the relative positional encoding a larger weight to assist the model in highlighting important positional pixels in the axial attention calculation, achieving accurate positioning of subtle lesions in the breast cancer segmentation task.

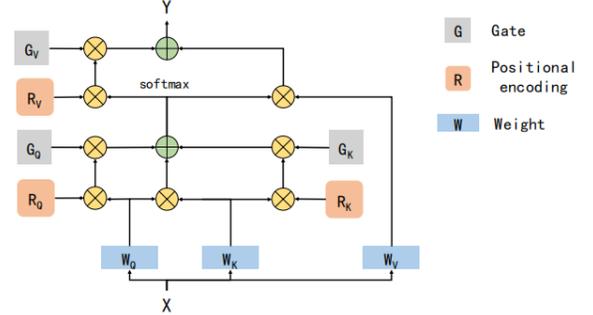

Figure 4. Schematic Diagram of the Gated Axial Attention Mechanism

## V. EXPERIMENT

### A. Dataset

The Breast Ultrasound Images Dataset (BUSI) [15] is a public dataset that was collected in 2018. It includes breast ultrasound images from women aged between 25 and 75 years old. The BUSI dataset comprises 780 breast ultrasound images from 600 patients, captured using the LOGIQ E9 ultrasound system. The average size of the images in BUSI is $500 \times 500$ pixels, and the format of the images is NG.

### B. Experiment Results Analysis

Table 1. Experiment Results

| Model | IoU | Precision | Recall | F1 |
|---|---|---|---|---|
| Segnet[16] | 0.7402 | 0.8061 | 0.8112 | 0.8086 |
| U-Net | 0.7731 | 0.8278 | 0.8371 | 0.8324 |
| Res-UNet[17] | 0.7795 | 0.8369 | 0.8554 | 0.8460 |
| Axial U-Net[18] | 0.7691 | 0.8577 | 0.8311 | 0.8441 |
| Axial Transformer | 0.7634 | 0.8813 | 0.8217 | 0.8504 |

As indicated in Table 1, the experiments quantitatively analyzed different models using precision, recall, intersection-over-union (IoU), and F1 score. In the BMRI dataset, the challenge of locating small breast cancer lesions is more pronounced than in the breast ultrasound image dataset due to the small size of the lesions. Analysis of the contents of Table 1 reveals that the proposed method surpasses mainstream convolutional neural network methods in the F1 score metric.

This suggests that the Axial Transformer can better capture long-term dependencies compared to other convolution-based segmentation models, mainly thanks to the multi-head attention mechanism of the Transformer. With better modeling of long-range dependencies, even when the background is much larger than the target, the Axial Transformer is less likely to misjudge the background as a lesion. This indicates that the Axial Transformer has indeed addressed the issue of difficult small lesion localization in breast cancer segmentation tasks.

In summary, the proposed Axial Transformer leverages the advantages of the Transformer model in modeling long-range dependencies for breast cancer segmentation tasks, combining axial attention and relative position encoding to achieve accurate localization of small breast cancer lesions. In performance comparisons, the segmentation precision and F1 score are higher than those of CNN segmentation models. However, the Axial Transformer has lower recall and IoU evaluation metrics, indicating a weaker ability to find all lesions. Radiologists have pointed out that, in addition to accurately locating small lesions, breast cancer segmentation models also need to accurately segment the blurred edges of lesions.

VI. CONCLUSIONS

In this paper, we introduced a transformative approach to breast cancer image segmentation by developing a novel Transformer-based model that employs an axial attention mechanism to overcome the limitations of traditional convolutional neural networks, such as U-Net, in accurately localizing small lesions. The proposed model decomposes the conventional self-attention mechanism into two one-dimensional axial attention along the horizontal and vertical axes, significantly enhancing the model's ability to capture critical features across the entire image while simultaneously reducing computational complexity. Recognizing the challenges posed by small datasets in medical imaging, we further incorporated a gated position-sensitive axial attention mechanism that leverages relative position information, allowing the model to focus more precisely on relevant lesions. This gated mechanism enhances the model's robustness and adaptability, particularly in scenarios with limited data, by regulating the attention focus, thereby reducing the impact of irrelevant background noise and occlusions. Our work not only advances the accuracy of breast cancer lesion segmentation but also provides a computationally efficient solution that can be seamlessly integrated into clinical workflows, offering a significant leap forward in computer-aided diagnostic tools and potentially improving early detection and treatment outcomes for breast cancer patients.